\documentclass[prb,preprint,showpacs,showkeys,preprintnumbers,amsmath,amssymb,superscriptaddress]{revtex4-1}
\usepackage{graphicx,color}

\begin{document}

\title{Effect of Gd doping and O deficiency on the Curie temperature of EuO}

\author{Nuttachai Jutong}
\email{nuttachai.jutong@physik.uni-giessen.de}
\altaffiliation[present address: ]{Physikalisch-Chemisches Institut, Justus-Liebig-Universit\"at Gie{\ss}en,
35392 Gie{\ss}en, Germany}
\affiliation{Institut f\"ur Physik, Universit\"at Augsburg, 86135 Augsburg, Germany}

\author{Thomas Mairoser}
\affiliation{Zentrum f\"ur elektronische Korrelationen und Magnetismus, Universit\"at Augsburg, 86159 Augsburg, Germany}

\author{Ulrich Eckern}
\email{ulrich.eckern@physik.uni-augsburg.de}
\affiliation{Institut f\"ur Physik, Universit\"at Augsburg, 86135 Augsburg, Germany}

\author{Udo Schwingenschl\"ogl}
\email{udo.schwingenschlogl@kaust.edu.sa}
\affiliation{KAUST, PSE Division, Thuwal 23955-6900, Kingdom of Saudi Arabia}


\begin{abstract}
The effect of Gd doping and O deficiency on the electronic structure, exchange interaction,
and Curie temperature of EuO in the cubic and tetragonal phases is studied by means of density
functional theory. For both defects, the Curie temperature is found to exhibit a distinct
maximum as a function of the defect concentration. The
existence of optimal defect concentrations is explained by the interplay of the on-site,
RKKY, and superexchange contributions to the magnetism.
\end{abstract}



\maketitle

\section{Introduction}

In recent years europium monoxide, EuO, has received considerable attention as a potential
material for spintronics, because of its special electronic and magnetic properties.
The compound has a rock salt structure, and it is a ferromagnetic insulator below
the Curie temperature of $T_C=69$ K.\cite{A. Mauger,Steeneken-2002} The
divalent Eu ions possess a large magnetic moment of 7 $\mu_B$, originating from the
half-filled 4$f$ states, which are separated by an energy gap of 1.12 eV from the Eu 5$d$
conduction band.\cite{thomas1} EuO is suitable as spin filter due to
its spin polarization of almost 100\%, as demonstrated both by
experiment\cite{Steeneken-2002,J.S.Moodera-2007,Schmehl-2007,thomas2} and theory.\cite{r1} Spin
filter tunneling junctions (metal/EuO/metal heterojunctions) based on polycrystalline
EuO have been studied in various experiments.\cite{J.S.Moodera-2004,E.Negusse-J.Appl-2006,
SANTOS-2008,Watson-2008,E.Negusse-J.Appl-2009,Martina-J.Appl-2009,Martina-Europhys-2009}
Integration of EuO on semiconducting GaAs,\cite{Swartz-2010} GaN,\cite{Schmehl-2007}
and Si\cite{Schmehl-2007,Caspers-2011} has been demonstrated. Particularly,
the possibility of growth on graphene \cite{Kawakami-2012} and topological insulators
\cite{Moodera-2013} is interesting for spintronics devices.

The ferromagnetism of EuO, in general, originates from indirect exchange, $J_1$, and
superexchange, $J_2$. It is widely accepted that the indirect exchange is governed
by the Eu 4$f$ and 5$d$ orbitals,\cite{ingle08,An-2013} whereas the superexchange
involves the hybridized Eu 4$f$ and O 2$p$ orbitals,\cite{Savrasov-2011} where
mediation by 6$s$ and 5$d$ states appears to be important.\cite{Dowben-2012,Miyazaki-2009}
It has been suggested by Ingle and Elfimov \cite{ingle08} that $T_C$ can be enhanced most
effectively by reducing the gap between the Eu 4$f$ and 5$d$ states, and by minimizing the
hybridization between the Eu 4$f$ and O 2$p$ states. In this context, rare earth doping 
with La, Lu, and Gd, has been studied experimentally,
\cite{Mairoser-2010, Sutarto-2009, Mairoser-2011, Shai-2012, Altendorf-2012, X.Wang-2010,
J.A.C.Santana1-2012, J.A.C.Santana2-2012, Melville-2012, Mairoser-2013, Melville1-2013}
by model approaches,\cite{A. Mauger, Arnold-2008, Burg-2012, Takahashi-2012} and by
first-principles calculations.\cite{Miyazaki-2010,ingle08, An-2013, Savrasov-2011,
J.A.C.Santana1-2012, Shai-2012, H.Wang-2012, J.K.Glasbrenner-2012} The effects of rare earth
doping, which is efficient only for low dopant concentrations, have been explained by
modifications of the on-site and RKKY interactions.\cite{An-2013} On the other hand,
enhancement of $T_C$ can also be achieved in O deficient EuO,\cite{Barbagallo-2010,
Barbagallo-2011,Monteiro-2013} and by the application of tensile strain.\cite{Dowben-2012} 
In fact, epitaxial growth of EuO on appropriate substrates can result in a tetragonal or
an orthorhombic structure.\cite{Melville1-2013}

Commonly used methods for modeling doping effects on the electronic structure of
EuO are the virtual crystal approximation\cite{J.A.C.Santana1-2012, Miyazaki-2010} and the
rigid band approximation.\cite{Shai-2012} First-principles calculations for Gd-doped EuO
by the supercell approach (partial substitution of Eu by Gd) have been reported in Ref.\
\onlinecite{H.Wang-2012} without addressing the exchange interaction. Insight into the
magnetism has been accomplished in Refs.\ \onlinecite{J.K.Glasbrenner-2012, An-2013} for
a restricted set of configurations.

In contrast, the purpose of our work is to investigate
the effect of Gd doping and O deficiency in the entire concentration range relevant for
experiment, focusing on the electronic structure, exchange interaction,
and $T_C$. We will start by introducing our methodology in Sect.\ \ref{sec:methods}, then
analyze first the effect of Gd doping (Sect.\ \ref{sec:Gd-doping}), and afterwards 
(Sect.\ \ref{sec:O-deficiency}) that of O deficiency. The conclusions are given in
Sect.\ \ref{sec:conclusion}.

\section{Methods}
\label{sec:methods}

In our first-principles calculations we use a linear combination of atomic
orbitals and Troullier-Martins norm-conserving relativistic pseudopotentials
(as implemented in the SIESTA code).\cite{siesta}
The wave functions are expanded in a $\zeta$+polarization basis, except for the
Eu 4$f$ states for which we use a single-$\zeta$ basis. A cutoff of 600 Ry is employed
together with $4\times4\times4$ and $6\times6\times4$ uniform meshes, respectively,
for sampling the Brillouin zones of the cubic and tetragonal phases. To achieve an
accurate description of the EuO band gap, we use the local density approximation with
on-site Coulomb repulsions ($U's$), and exchange parameters ($J's$).\cite{anis91,anismov93}
Note that these parameters refer to the microscopic interacting-electron problem; in
particular, these $J's$ should not be confused with the exchange interactions of the
effective Heisenberg model, to be discussed at the end of this section.

For the Eu 4$f$ states we set $J_f=0.77$ eV,\cite{ingle08} but
we vary $U_f$ between 8 and 9 eV since the band structure
depends critically on the on-site potential of the Eu 4$f$ states. A value of $U_f=8.8$
eV gives the best agreement with the experimental situation (band gap of 1.1 eV, and band
splitting of 0.6 eV).\cite{J.S.Moodera-2007,Steeneken-2002} Following
Ref.\ \onlinecite{ingle08} we use $J_p=1.2$ eV and $U_p=4.6$ eV for the O 2$p$ orbitals.
Moreover, for the Gd 4$f$ orbitals we set $J_f=0.7$ eV and $U_f=6.7$ eV.\cite{anismov97,wang-2012}
  
\begin{figure}
\includegraphics[width=0.25\textwidth,clip=true]{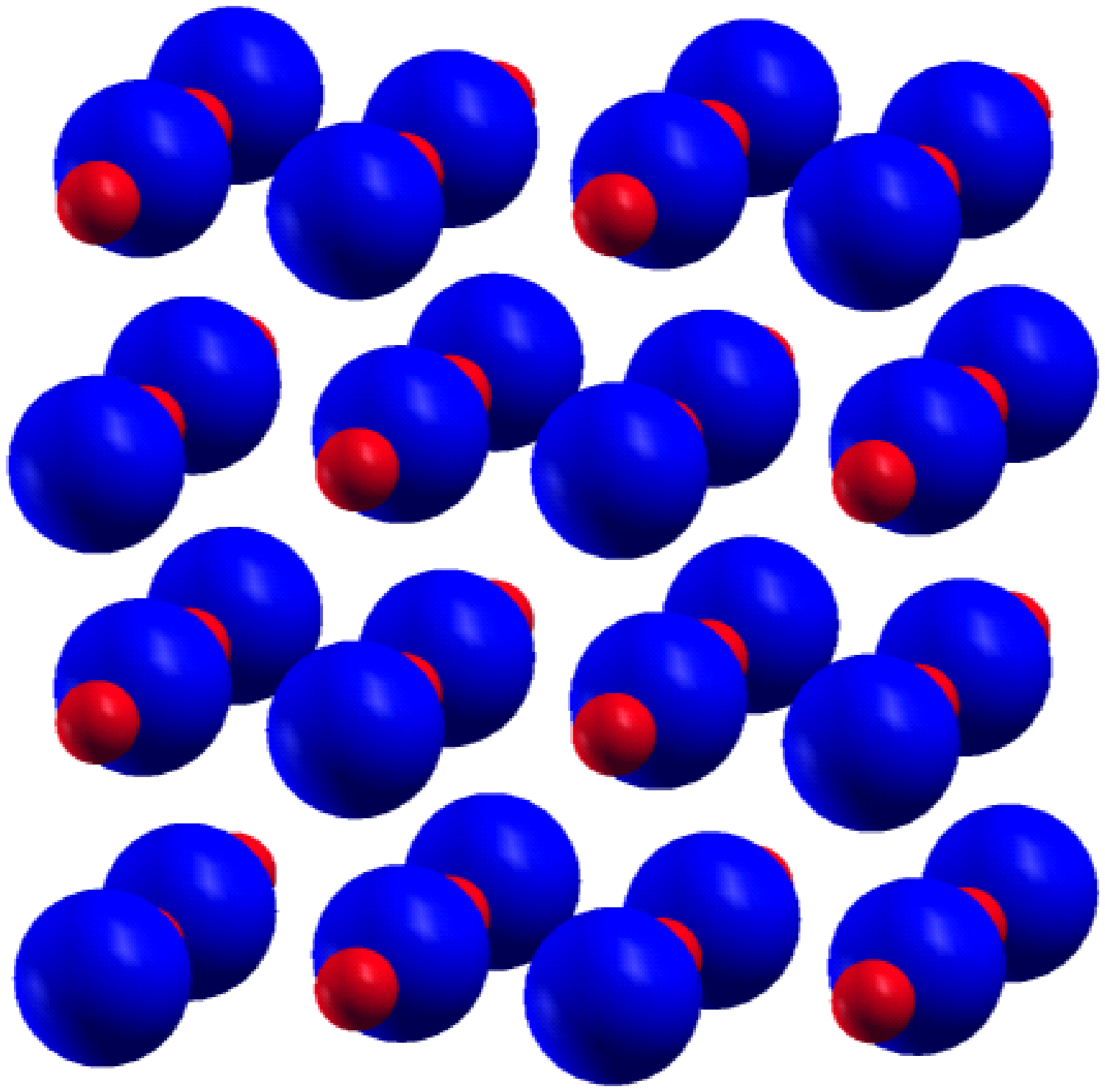}\hspace{0.5cm}
\includegraphics[width=0.2\textwidth,clip=true]{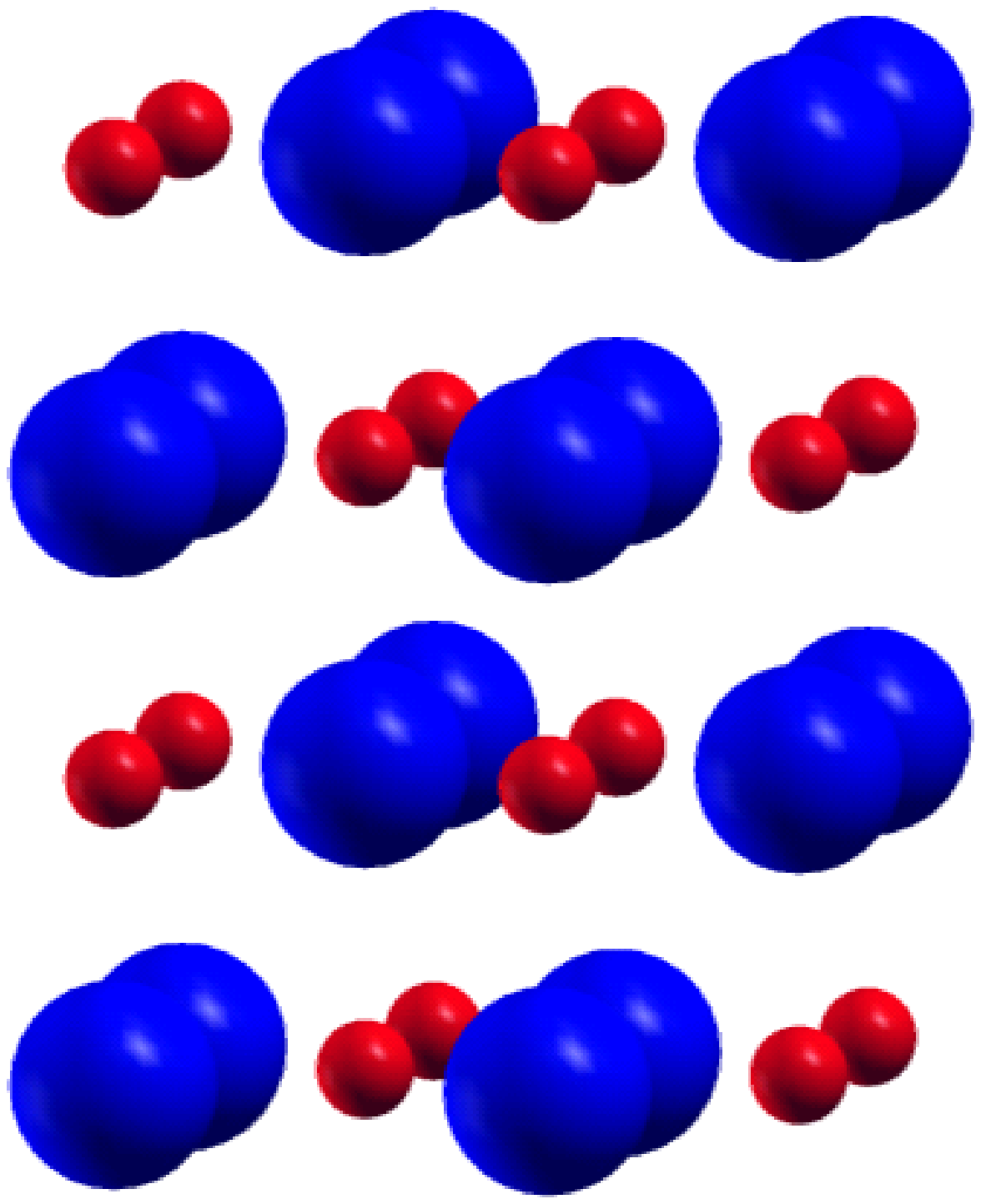}
\caption{(Color online) Structure of EuO in the cubic phase (left) and the tetragonal phase
(right). Large spheres (blue) represent Eu, small spheres (red) represent O. The front side
is the $xz$-plane.}
\label{Fig1}
\end{figure}

The experimental lattice constant of 5.144 \AA\ is used for the rocksalt structure (cubic
phase), with four Eu and four O atoms per unit cell. For the tetragonal phase we start from the
lattice parameters $a=3.65$ \AA\ (inter-planar spacing) and $c=5.12$ \AA\ (out-of-plane spacing).
For both phases, $2\times2\times2$ supercells are built, which
are shown in Fig.\ \ref{Fig1}. These supercells are fully relaxed by means of the conjugate
gradient method until the atomic forces have declined below 0.01 eV/\AA. We obtain for
the cubic phase $a=5.097$ \AA\, and for the tetragonal phase $a=3.635$ \AA\ and
$c=5.080$ \AA. These values are kept fixed when building the respective structures under
Gd doping and O deficiency. Under this constraint, we have carefully relaxed all atomic
positions which, of course, is mandatory in order to be able to obtain reliable results.
Gd concentrations between 6.25\% and 25\% are considered,
by substituting Eu atoms by Gd; O vacancy concentrations in the same range are
achieved by removing O atoms from the supercell.

The nearest neighbor (NN, $J_1$) and next-nearest neighbor (NNN, $J_2$) exchange
interactions for the cubic phase are determined by fixing three spin configurations,
and calculating their respective energies:
the ferromagnetic one (FM), an antiferromagnetic (AFM) one with the spin direction alternating
in the (001) direction (AFM1), and an AFM one with the spin direction changing every second layer in
the (001) direction (AFM2). The total energies (per cation) are related to the $J_1$ and $J_2$
as follows:\cite{Larson-2006} 
\begin{eqnarray}
&&E_{\rm FM} = E_0+S(S+1)(-12J_1-6J_2),\\\nonumber
&&E_{\rm AFM1} = E_0+S(S+1)(4J_1-6J_2),\\\nonumber
&&E_{\rm AFM2} = E_0+S(S+1)(-4J_1-2J_2)
\end{eqnarray}
where $S=7/2$. Given $J_1$ and $J_2$, an effective Heisenberg model can be defined; and
similarly for the tetragonal case (next paragraph).

For the tetragonal phase ($c>a$), we have in-plane NN ($J_{1\parallel}$),
out-of-plane NN ($J_{1\perp}$), in-plane NNN ($J_{2\parallel}$), and out-of-plane NNN
($J_{2\perp}$) interactions. Note that the term in-plane refers to the $xy$-plane of
the tetragonal supercell, which is rotated by 45$^\circ$ with respect to the cubic
supercell. To determine the exchange interactions in this case, 
we have to study five spin configurations:
FM, AFM1, AFM2, AFM with the spin direction alternating in the (110) direction
(AFM3), and AFM with the spin direction alternating every 2nd layer in the (100)
direction (AFM4). The total energies (per cation) are given by:
\begin{eqnarray}
&&E_{\rm FM} = E_0+S(S+1)(-4J_{1\parallel}-8J_{1\perp}-4J_{2\parallel}-2J_{2\perp}),\\\nonumber
&&E_{\rm AFM1} = E_0+S(S+1)(-4J_{1\parallel}+8J_{1\perp}-4J_{2\parallel}-2J_{2\perp}),\\\nonumber
&&E_{\rm AFM2} = E_0+S(S+1)(-4J_{1\parallel}-4J_{2\parallel}+2J_{2\perp}),\\\nonumber
&&E_{\rm AFM3} = E_0+S(S+1)(4J_{1\parallel}-4J_{2\parallel}-2J_{2\perp}),\\\nonumber
&&E_{\rm AFM4} = E_0+S(S+1)(4J_{2\parallel}-2J_{2\perp}).
\end{eqnarray}        
In mean-field approximation this results in\cite{ingle08}
\begin{eqnarray}
&&T_C^{\rm cubic}=\frac{2}{3}S(S+1)(12J_1+6J_2),\\\nonumber
&&T_C^{\rm tetra}=\frac{2}{3}S(S+1)(4J_{1\parallel}+8J_{1\perp}+4J_{2\parallel}+2J_{2\perp}).
\end{eqnarray}

\section{Gd doping}
\label{sec:Gd-doping}

\begin{figure}
\includegraphics[width=0.5\textwidth,clip=true]{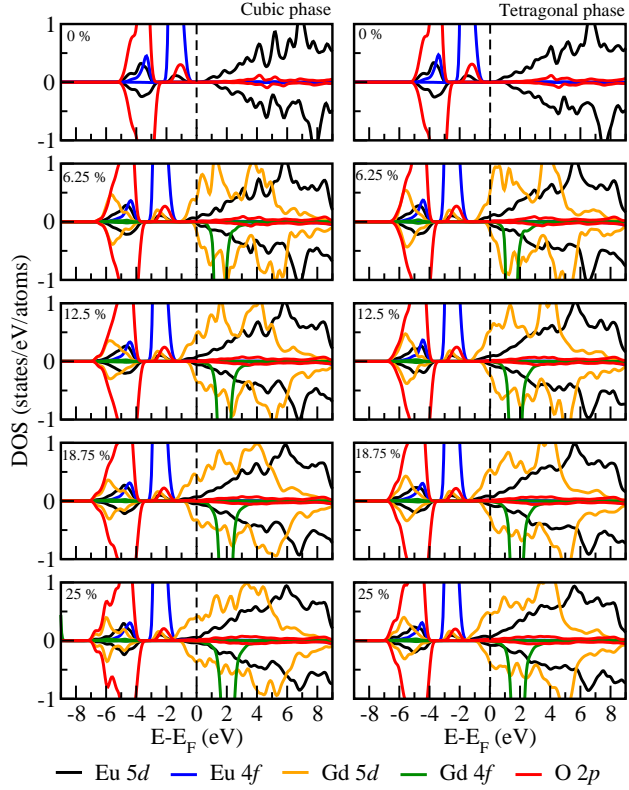}
\caption{DOS projected on the Eu 4$f$, 5$d$, Gd 4$f$, 5$d$, and O 2$p$ orbitals
for the cubic (left) and tetragonal (right) phases of EuO for different Gd concentrations}
\label{Fig2GdPDOSRx}
\end{figure}

\begin{figure}
\includegraphics[width=0.5\textwidth,clip=true]{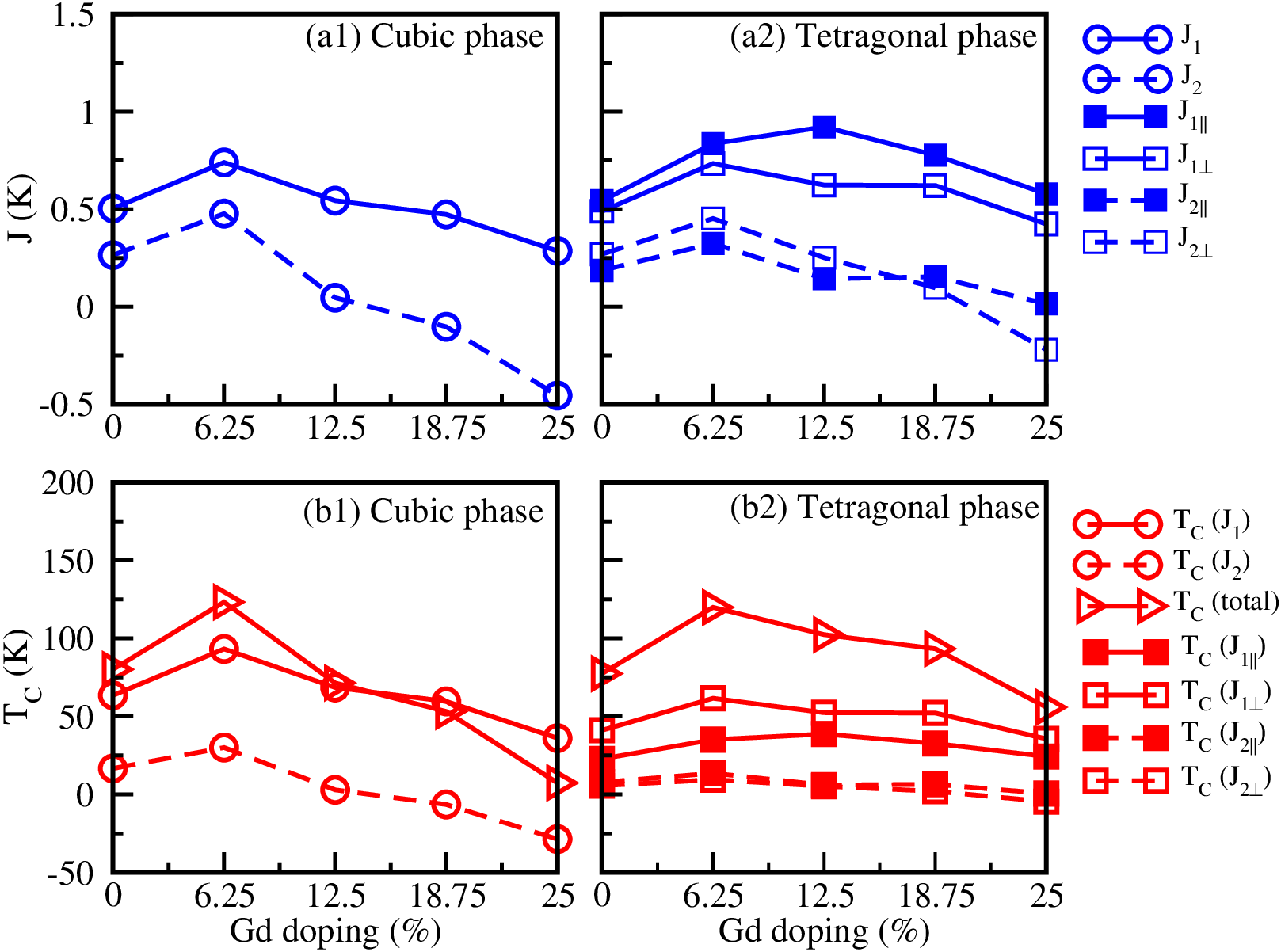}
\caption{Exchange interaction and corresponding $T_C$ as a function of the
Gd concentration for the cubic phase ((a1), (b1)), and the tetragonal phase ((a2), (b2))}   
\label{Fig3-Compare-JTc-GdDoping}
\end{figure}

In order to clarify the effect of Gd doping for both
the cubic and tetragonal phases we determine the density of states (DOS) projected
on the Eu 4$f$, 5$d$, Gd 4$f$, 5$d$, and O 2$p$ orbitals, see Fig.\ \ref{Fig2GdPDOSRx}. The
dependences of the different exchange terms and of $T_C$ on the dopant concentration are
addressed in Fig.\ \ref{Fig3-Compare-JTc-GdDoping}. We first discuss the results for the
pristine structures (0\% doping), which are very similar for the cubic and tetragonal phases.
For the majority spin channel we distinguish three regions: the conduction band (dominated by
Eu 5$d$ states), upper valence band (dominated by localized Eu 4$f$ states, with some
hybridization with O 2$p$ and Eu 5$d$), and lower valence band (dominated by O 2$p$ states,
with significant hybridization with Eu 4$f$ and 5$d$). The spin minority channel shows
a similar structure but without the Eu 4$f$ contributions. The energy gap between the
valence and conduction bands amounts to 1.1 eV, and the exchange splitting of the Eu 5$d$ states
at the conduction band edge to 0.6 eV, in good agreement with the experiment.\cite{J.S.Moodera-2007}

For the cubic phase, exchange interactions of $J_1=0.63$ K and $J_2=0.13$ K have been
derived from single-crystal inelastic neutron scattering,\cite{Mook-1981} which
by Eq.\ (3) corresponds to $T_C=88$ K, while we find $J_1=0.50$ K and $J_2=0.26$ K and
hence a $T_C$ of 80 K, consistent with the experimental result. Note that our effective 
$J$ ($= J_1 + J_2 = 0.76$ K) agrees with the experimental value.
For the tetragonal phase, we obtain $J_{1\parallel} = 0.54$ K, $J_{2\parallel} = 0.19$ K, 
$J_{1\perp} = 0.49$ K, and $J_{2\perp} = 0.27$ K, from which a mean-field $T_C$ of 77 K
is calculated.
However, recent experiments on films of cubic and tetragonal EuO with 10 \AA\ thickness have
found critical temperatures of 56 K and 53 K, respectively.\cite{Melville1-2013} While the
absolute values deviate from our theoretical findings, we note that the difference
between the two $T_C$'s is exactly the same (3 K), This is a strong indication for the
reliability of our calculations, as far as difference quantities and dependencies (like
$T_C$ vs.\ concentration) are concerned.

The effects of Gd doping on the DOS are similar for the cubic and tetragonal phases,
see Fig.\ \ref{Fig2GdPDOSRx}. The exchange splitting of the Eu 5$d$ states at the conduction
band edge essentially remains the same as in the pristine system. For increasing Gd doping,
the Gd 5$d$ and Eu 5$d$ majority spin states shift to lower energy, increasing the system's
metallicity. Since there are many more Gd 5$d$ than Eu $5d$ conduction states
occupied, mainly the Gd 5$d$ states determine $J_1$ (combination of on-site and RKKY
exchange). The hybridization between the Eu 4$f$ and O 2$p$ states decreases for increasing
Gd doping, which reduces the value of $J_2$ (superexchange). 
 
For 6.25\% Gd doping the stronger exchange interaction between the Gd/Eu 5$d$
and Eu 4$f$ states (the reduced energy gap supports the $f$-$d$ hopping) in combination
with the RKKY exchange mediated by the conduction states\cite{An-2013} enhances
$T_C$ to around 120 K, both in the cubic and tetragonal phases, see
Fig.\ \ref{Fig3-Compare-JTc-GdDoping}(b1),(b2), in good agreement with the experimental
value of 129 K for 10\% Gd doping.\cite{Mairoser-2010} In Ref.\ \onlinecite{An-2013} a
maximal $T_C$ of 160 K for 10\% Gd doping has been obtained on the basis of the virtual
crystal approximation (using the parameters $J_f=0.6$ eV and $U_f=6.1$ eV for the Eu 4$f$
states). While the validity of the virtual crystal approximation is difficult to assess
in this context, we note that $T_C$ depends strongly on the parameter $U_f$, which in
our work was chosen to be $U_f=8.8$ eV, in order to reproduce the experimental band gap.
With increasing $U_f$, the band gap opens, hence $T_C$ decreases, and vice versa.

Above 6.25\% Gd doping we observe that the Gd/Eu 5$d$
majority spin states shift further towards the Eu 4$f$ states, which should enhance $T_C$.
However, the minority spin states start getting filled and, as a consequence, the spin
polarization at the Fermi energy is reduced. This compromises the RKKY interaction and
therefore effectively lowers $J_1$ and $T_C$. In addition, an antiferromagnetic $J_2$ is
observed for 18.75\% and higher doping in both phases, which can be explained by enhanced
hybridization between the Gd 5$d$, Eu 5$d$, and O $2p$ states:
see, for example, the developing joint DOS peaks close to $-5$ eV.

\section{O deficiency}
\label{sec:O-deficiency}

\begin{figure}
\includegraphics[width=0.5\textwidth,clip=true]{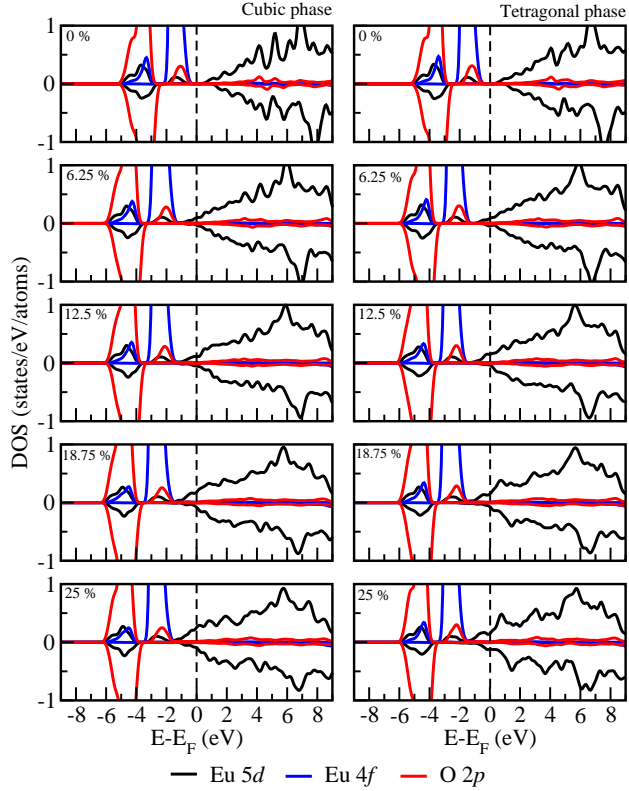}
\caption{DOS projected on the Eu 4$f$, Eu 5$d$, and O 2$p$ orbitals for the cubic (left)
and tetragonal (right) phases of EuO for different O vacancy concentrations}
\label{Fig3VoPDOSRx}
\end{figure}

\begin{figure}
\includegraphics[width=0.5\textwidth,clip=true]{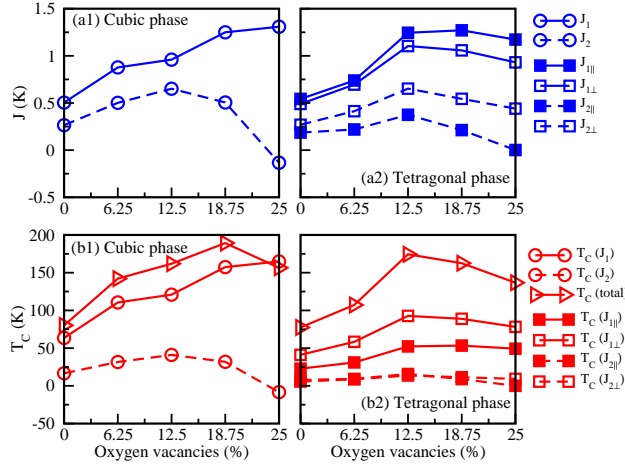}
\caption{Exchange interaction and corresponding $T_C$ as a function of the
O vacancy concentration for the cubic phase ((a1), (b1)) and the tetragonal phase ((a2), (b2))} 
\label{Fig41-Compare-J-Tc-Vo}
\end{figure}  

We next analyze the effects of O deficiency by means of the DOS projected on the Eu 4$f$, 5$d$
and O 2$p$ orbitals, see Fig.\ \ref{Fig3VoPDOSRx}, for different O vacancy concentrations.
In addition, Fig.\ \ref{Fig41-Compare-J-Tc-Vo} addresses the dependences of the different
exchange terms and of $T_C$ on the O vacancy concentration. As expected, O deficiency
causes almost rigid shifts of all states to lower energy, so that more and more of the
charge donated by the O vacancies occupies the Eu $5d$ conduction bands. It is generally
accepted that positive effects on $T_C$ due to O deficiency originate from this extra charge
populating the conduction band, and giving rise to enhanced RKKY
exchange,\cite{A. Mauger, Dowben-2012, Barbagallo-2010,Barbagallo-2011}
which corresponds to an increase in $J_1$. However, also the gap between the majority spin
Eu $5d$ and 4$f$ states decreases substantially, and the $f$-$d$ hopping
is enhanced correspondingly, see Fig.\ \ref{Fig41-Compare-J-Tc-Vo}(a1). The band
structure (not shown) demonstrates that the exchange splitting of the Eu 5$d$ states at
the conduction band edge is reduced significantly for 6.25\% O vacancy concentration,
as compared to the pristine case, and further slightly decreases for higher O vacancy
concentrations. 

In addition, the DOS demonstrates that hybridization between the Eu $4f$, $5d$ and O 2$p$
states plays a significant role for the $T_C$ value. We first focus on
the cubic phase, see the left hand side of Fig.\ \ref{Fig3VoPDOSRx}. Hybridization between
the Eu 4$f$ and O 2$p$ states decreases as the O vacancy concentration increases, which
enhances $J_2$, up to 12.5\% O vacancy concentration. Afterwards $J_2$ declines rapidly.
According to Fig.\ \ref{Fig41-Compare-J-Tc-Vo}(b1), $T_C$ increases, as $J_1$ increases,
up to a maximum value of 190 K for 18.75\% O vacancy concentration, and decreases thereafter,
as $J_2$ decreases. The tetragonal phase, see the right hand side of Fig.\ \ref{Fig3VoPDOSRx},
overall shows similar characteristics, i.e., $J_{1\parallel}$,
$J_{1\perp}$, $J_{2\parallel}$, and $J_{2\perp}$, see Fig.\ \ref{Fig41-Compare-J-Tc-Vo}(a2),
first increase with O deficiency. However, now the Eu $5d$ minority spin states get filled
for 12.5\% and higher O vacancy concentrations, and $J_1$ is reduced accordingly, resulting
in a maximum in $T_C$ of about 175 K, see Fig.\ \ref{Fig41-Compare-J-Tc-Vo}(b2).
    
\section{Conclusion}
\label{sec:conclusion}
We have performed first principles calculations for both Gd doped and O deficient EuO
to clarify the mechanisms that determine the critical temperatures of the cubic and
tetragonal phases. We extend previous theoretical considerations for the cubic phase to
high defect concentrations, and present the first comprehensive account of the role of
defects in the tetragonal phase. The calculated maximum in $T_C$, as a function of
Gd concentration, is in good agreement with the experimental value. The observed behavior
is explained by a complex combination of different exchange mechanisms. While both
the on-site and RKKY interactions increase with increasing (but low) doping,
filling of the Gd 5$d$ minority
spin states at high doping counteracts the RKKY exchange. In addition, the superexchange is
modified at high doping due to growing hybridization between the Gd 4$f$ and O 2$p$ states.
The dependence of $T_C$ on the O deficiency is controlled by a similar mechanism, though
now the Eu $5d$ states take over the role of the Gd $5d$ states. As a consequence,
optimal values exist both for the Gd dopant and O vacancy concentrations.

\section{Acknowledgements}
We thank T.\ Archer, L.\ Chioncel, and I. Rungger for fruitful discussions.
The work in Augsburg was supported by the Deutsche Forschungsgemeinschaft
(TRR 80). Computational resources have been provided by LRZ Munich, Germany.

\end{document}